\documentstyle[12pt]{article}
\begin{document}
\tolerance=5000
\def\be{\begin{equation}}
\def\ee{\end{equation}}
\def\bea{\begin{eqnarray}}
\def\eea{\end{eqnarray}}
\def\nn{\nonumber \\}
\def\cF{{\cal F}}
\def\det{{\rm det\,}}
\def\Tr{{\rm Tr\,}}
\def\e{{\rm e}}
\def\etal{{\it et al.}}
\def\erp2{{\rm e}^{2\rho}}
\def\erm2{{\rm e}^{-2\rho}}
\def\er4{{\rm e}^{4\rho}}
\def\etal{{\it et al.}}

\ 

\vskip -2cm

\ \hfill
\begin{minipage}{3.5cm}
NDA-FP-50 \\
October 1998 \\
\end{minipage}

\vfill

\begin{center}
{\Large\bf  Conformal Anomaly for Dilaton Coupled Theories 
from AdS/CFT Correspondence}

\vfill

{\sc Shin'ichi NOJIRI}\footnote{\scriptsize 
e-mail: nojiri@cc.nda.ac.jp} and
{\sc Sergei D. ODINTSOV$^{\spadesuit}$}\footnote{\scriptsize 
e-mail: odintsov@tspi.tomsk.su}

\vfill

{\sl Department of Mathematics and Physics \\
National Defence Academy, 
Hashirimizu Yokosuka 239, JAPAN}

\ 

{\sl $\spadesuit$ 
Tomsk Pedagogical University, 634041 Tomsk, RUSSIA \\
}

\ 

\vfill

{\bf abstract}

\end{center}

Trace anomaly for dilaton coupled conformal theories on
curved background with non-zero dilaton is found from supergravity side 
as an IR effect using AdS/CFT correspondence. For $d=2$ it coincides 
with the conformal anomaly for dilaton coupled scalar 
(up to total derivative term 
which is known to be ambiguous). In four-dimensional case we get conformal 
anomaly for ${\cal N}=4$ super YM theory interacting with 
conformal supergravity. In the same way the calculation of dilaton 
dependent conformal anomaly in higher dimensions seems to be much easier 
than using standard QFT methods.

\ 


\newpage

AdS/CFT correspondence \cite{M,W,GKP} attracted a lot of attention 
recently. This correspondence between string theory on specific 
background and conformal field theory is simplified for large $N$ (where 
$N$ is the number of coincident branes) and small string coupling. 
In this case, the left hand side of conjecture relation, i.e. string 
partition function in supergravity approximation may be reduced to 
exponential of the supergravity action functional which is calculated on 
the background under consideration (bulk) with the account of 
classical equations of motion and some boundary conditions. The right 
hand side is given by generating functional of correlation functions in 
the conformal field theory on the boundary. This partition function may 
be also considered as the one corresponding to the coupling of 
conformal matter with conformal supergravity \cite{HT}.

Recently conformal anomaly (for a general review and list of
refs., see \cite{MD}) for conformal field theories in $d=2,4,6$ has 
been derived from the supergravity side using above AdS/CFT correspondence 
in ref.\cite{HS}. Such calculation explicitly demonstrates that 
conformal anomaly which usually has an UV origin as it is almost 
associated with the corresponding divergent effective action 
(see \cite{BOS} for a review) may arise as an IR divergence in bulk theory 
(so-called IR-UV connection in holographic theories \cite{SW}). 

From another side recently conformal anomaly has been calculated by
usual QFT methods for dilaton coupled conformal matter on curved background 
with non-trivial dilaton. For 2d dilaton coupled scalar with arbitrary
dilatonic coupling it has been first calculated in ref.\cite{E} (see 
also \cite{NO} and for special choice of dilatonic coupling, 
see\cite{BH}). For 4d dilaton coupled scalars and vectors the 
conformal anomaly has been found in refs.\cite{NO,NO1,IO}.

The main purpose of our work is to find holografic conformal anomaly 
for dilaton coupled conformal field theory from supergravity side (using 
AdS/CFT correspondence). We use the same methods as in ref.\cite{HS} 
where such result has been obtained in the absence of background dilaton.
The final expression for conformal anomaly is given for $d=2$ and $d=4$. 
It is shown that for $d=2$ it coincides with the conformal anomaly for 
2d dilaton coupled scalar \cite{E,NO} and it is actually unique (up 
to total derivatives terms \cite{NO,BH}). For $d=4$ we get just 
conformal anomaly of ${\cal N}=4$ $U(N)$ or $SU(N)$ super YM theory 
coupled with ${\cal N}=4$ conformal supergravity.

We start with the action of $d+1$-dimensional dilatonic gravity
\bea
\label{i}
S&=&{1 \over 16\pi G}\left[\int_{M_{d+1}} d^{d+1}x \sqrt{-\hat G}
\left\{ \hat R + X(\phi)(\hat\nabla\phi)^2 + Y(\phi)\hat\Delta\phi
+ 4\lambda^2  \right\}\right. \nn
&& \left. +\int_{M_d} d^dx\sqrt{-\hat g}
\left(2\hat\nabla_\mu n^\mu + \alpha\right)  \right]\ .
\eea
Here $M_{d+1}$ is the $d+1$ dimensional manifold whose boundary is the 
$d$ dimensional manifold $M_d$ and $n_\mu$ is the unit normal vector 
to $M_d$. In Eq.(\ref{i}) $X(\phi)$ and $Y(\phi)$ are arbitrary functions 
depending on dilaton $\phi$. Note that the arbitrariness of $X(\phi)$ 
and $Y(\phi)$ is not real but apparent one. 
In fact, if we define new dilaton field $\varphi$ by
\be
\label{xviii}
\varphi\equiv\int d\phi\sqrt{2V(\phi)}\ ,\ \ 
V(\phi)\equiv X(\phi)-Y'(\phi)\ ,
\ee
the action (\ref{i}) can be rewritten by using partial integration 
as follows:
\bea
\label{ib}
S&=&{1 \over 16\pi G}\left[\int d^{d+1}x \sqrt{-\hat G}
\left\{ \hat R + 2(\hat\nabla\varphi)^2 + 4\lambda^2  \right\}\right. \nn
&& \left. +\int_{M_d} d^dx\sqrt{-\hat g}
\left(2\hat\nabla_\mu n^\mu 
+ \alpha + Y(\phi)n^\mu\partial_\mu\phi \right) \right]\ .
\eea
The $\phi$ dependent term on $M_d$ does not finally contribute to 
Weyl anomaly. 
We keep, however, $X(\phi)$ and $Y(\phi)$ as arbitrary functions for 
the later convenience. Note also that boundary term may be used to 
present the action as the functional of fields and their first 
derivatives \cite{GH}.

As in \cite{HS}, we choose the metric $\hat G_{\mu\nu}$ on $M_{d+1}$ and 
the metric $\hat g_{\mu\nu}$ on $M_d$ in the following form (see also
\cite{FG}):
\bea
\label{ii}
ds^2&\equiv&\hat G_{\mu\nu}dx^\mu dx^\nu 
= {l^2 \over 4}\rho^{-2}d\rho d\rho + \sum_{i=1}^d
\hat g_{ij}dx^i dx^j \nn
\hat g_{ij}&=&\rho^{-1}g_{ij}\ .
\eea
Here $l$ is related with $\lambda^2$ by $4\lambda^2=-d(d-1)/l^2$.
Note that the expression of the metric (\ref{ii}) has a redundancy.
In fact, the expression (\ref{ii}) is invariant if we change $\rho$ 
and $g_{ij}$ by 
\be
\label{wtr}
\delta\rho= \delta\sigma\rho\ ,\ \ 
\delta g_{ij}= \delta\sigma g_{ij}\ .
\ee
Here $\delta\sigma$ is a constant parameter of the transformation.
The transformation (\ref{wtr}) can be regarded as the scale 
transformation on $M_d$.

Using the expression  (\ref{ii}), we find that the unit vector $n_\mu$ 
normal to $M_d$ can be given by $\left(0,0,\cdots,0,{2\rho \over l}\right)$ 
and we find
\be
\label{iia}
\hat\nabla_\mu n^\mu =-{d \over l}+{\rho \over l}g^{ij}g'_{ij}\ .
\ee
Here ``$\ '\ $" expresses the derivative with respect to $\rho$. 
We also find that the scalar curvature $\hat R$ has the following form
\bea
\label{iii}
\hat R&=&{d^2 + d \over l^2}+\rho R \nn
&&-{2(d-1) \rho \over l^2}g^{ij}g'_{ij}
-{3\rho^2 \over l^2} g^{ij}g^{kl}g'_{ik}g'_{jl}
+ {4\rho^2 \over l^2}g^{ij}g''_{ij}
+{\rho^2 \over l^2}g^{ij}g^{kl}g'_{ij}g'_{kl}
\eea
Here $R$ is the scalar curvature on $M_d$.

When $d$ is even, we can expand $\phi$ and $\rho$ 
as power series of $\rho$ as in \cite{HS}:
\bea
\label{iv}
\phi&=&\phi_{(0)}+\rho\phi_{(1)}+\rho^2\phi_{(2)}
+\cdots \rho^{d \over 2}\phi_{(d/2)}
- \rho^{d \over 2}\ln\rho \psi + {\cal O}(\rho^{{d \over 2}+1}) \\
g_{ij}&=&g_{(0)ij}+\rho g_{(1)ij}+\rho^2 g_{(2)ij}+\cdots 
+\rho^{d \over 2}g_{(d/2)ij}-\rho^{d \over 2}\ln\rho h_{ij}
+ {\cal O}(\rho^{{d \over 2}+1}) \ . \nonumber
\eea
Here we regard $\phi_{(0)}$ and $g_{(0)ij}$ as independent fields on 
$M_d$ and $\phi_{(l)}$, $g_{(l)ij}$ ($l=1,2,\cdots$), $\psi$ and $h_{ij}$ 
as fields depending on $\phi_{(0)}$ and $g_{(0)ij}$ by using equations of 
motion. Then the action (\ref{i}) diverges in general since the action 
contains the infinite volume integration on $M_{d+1}$. 
The action is regularized by introducing the infrared cutoff $\epsilon$ 
and replacing 
\be
\label{vi}
\int d^{d+1}x\rightarrow \int d^dx\int_\epsilon d\rho \ ,\ \ 
\int_{M_d} d^d x\Bigl(\cdots\Bigr)\rightarrow 
\int d^d x\left.\Bigl(\cdots\Bigr)\right|_{\rho=\epsilon}\ .
\ee
As discussed in \cite{HS}, the terms proportional to the (inverse) power of 
$\epsilon$ in the regularized action are invariant under the scale 
transformation 
\be
\label{via}
\delta g_{(0)\mu\nu}=2\delta\sigma g_{(0)\mu\nu}\ ,\ \  
\delta\epsilon=2\delta\sigma\epsilon \ , 
\ee
which corresponds to (\ref{wtr}).  
Then the subtraction of these terms proportional to the inverse power of 
$\epsilon$ does not break the invariance under the scale transformation. 
When $d$ is even, however, the term proportional to $\ln\epsilon$ appears.
The term is not invariant under the scale transformation (\ref{via}) and the 
subtraction of the $\ln\epsilon$ term breaks the invariance. 
The variation of the $\ln\epsilon$ term under the scale 
transformation (\ref{via}) is finite when $\epsilon\rightarrow 0$ and 
should be canceled by the variation of the finite term (which does not 
depend on $\epsilon$) in the action since the original action (\ref{i}) 
is invariant under the scale transformation. 
Therefore the $\ln\epsilon$ term $S_{\rm ln}$ gives the Weyl anomaly $T$ 
of the action renormalized by the subtraction of the terms which diverge 
when $\epsilon\rightarrow 0$ by \cite{HS}
\be
\label{vib}
S_{\rm ln}=-{1 \over 2}
\int d^4x \sqrt{-g_{(0)}}T\ .
\ee

First we consider the case of $d=2$. Choosing $\alpha$ to satisfy the 
equation $\alpha={2 \over l}$ and  using the replacement in (\ref{vi}), 
we find the action (\ref{i}) has the following form
\bea
\label{vii}
S&=&-{1 \over 16\pi G}{l \over 2}\ln\epsilon\int d^2x
\sqrt{-g_{(0)}}
\left\{ R_{(0)}+ X(\phi_{(0)})(\nabla\phi_{(0)})^2 
\right.\nn
&&\left. + Y(\phi_{(0)})\Delta\phi_{(0)}
\right\} 
+\ \mbox{finite terms}\ .
\eea
Then we find an expression of the Weyl anomaly $T$ by using (\ref{vib})
\be
\label{ix}
T={l \over 16\pi G} \left\{ R_{(0)}
+ X(\phi_{(0)})(\nabla\phi_{(0)})^2 
+ Y(\phi_{(0)})\Delta\phi_{(0)}
\right\}\ .
\ee
The trace anomaly for $N$ dilaton coupled  matter scalar 
fields whose action is given by
\be
\label{x}
S={1 \over 2}\int d^2x \sqrt{-g}
f(\phi) g^{\mu\nu}\sum_{i=1}^N\partial_\mu\chi_i
\partial_\nu\chi_i
\ee
has been calculated in Ref.\cite{E,NO}. The result corresponds to
\bea
\label{xi}
&& {l \over 16\pi G} ={N \over 24\pi} \ ,\ \ 
{l \over 16\pi G}X(\phi_{(0)})=-{N \over 4\pi}
\left({f'' \over 2f}- {{f'}^2 \over 4f^2}\right) \ ,\nn
&& {l \over 16\pi G}Y(\phi_{(0)})=-{N \over 4\pi}{f' \over 2f}
\ .
\eea
The above result should give the conformal anomaly computed from 
the asymptotic symmetry algebra of AdS$_3$ with dilaton, as it was the case 
in the absence of dilaton \cite{JBH}. Substituting (\ref{xi}) into 
(\ref{i}) and integrating it by parts, we find
\bea
\label{xii}
S&=&{N \over 24\pi l}\left[\int d^{d+1}x \sqrt{-\hat G}
\left\{\hat R - 6(\hat\nabla\tilde\phi)^2 
+ 4\lambda^2 \right\} \right.\nn
&& \left. 
+\int_{M_d} d^dx\sqrt{-\hat g}
\left(2\hat\nabla_\mu n^\mu + \alpha + 6n^\mu \hat\nabla_\mu\tilde\phi
\right) \right]\ .
\eea
Here $\tilde\phi=-{1 \over 2}\ln f(\phi)$.
Note that we can forget the classical kinetic term of the dilaton
(with gravity) when we only consider the anomaly from scalars 
coupled with dilaton as in (\ref{x}), for example, in case of large $N$. 
Therefore we can freely redefine the dilaton field, say we can 
regard  $\tilde\phi=-{1 \over 2}\ln f(\phi)$ with the redefined dilaton 
field for the arbitrary (positive) dilaton function $f(\phi)$. 
Then with these redefinitions, the resulting anomaly is almost unique 
although it is still general, up to total derivative terms.   

We now consider the case of 4 dimensions. 
Tedious calculation similar to that in 2 dimensions leads to the 
term $S_{\rm ln}$ proportional to $\ln\epsilon$ in the action in 
the following form:
\bea
\label{xiii}
S_{\rm ln}&=&{1 \over 16\pi G}\int d^4x \sqrt{-g_{(0)}}\left[ 
{1 \over 2l}g_{(0)}^{ij}g_{(0)}^{kl}\left(g_{(0)ij}g_{(0)kl}
-g_{(0)ik}g_{(0)jl}\right) \right. \nn
&& +{l \over 2}\left(R_{(0)}^{ij}-{1 \over
2}g_{(0)}^{ij}R_{(0)}\right)g_{(1)ij} \nn
&& -{2 \over l}\left(X(\phi_{(0)})-Y'(\phi_{(0)})\right)\phi_{(1)}\phi_{(1)}
-{l \over 2}X'(\phi_{(0)})\phi_{(1)}
g_{(0)}^{ij}\partial_i\phi_{(0)}\partial_j\phi_{(0)} \nn
&& -l X(\phi_{(0)})g_{(0)}^{ij}\partial_i\phi_{(0)}\partial_j\phi_{(1)} 
+{l \over 2}X(\phi_{(0)})g_{(0)}^{ik}g_{(0)}^{jl}
g_{(1)kl}\partial_i\phi_{(0)}\partial_j\phi_{(0)} \nn
&& -{l \over 4}X(\phi_{(0)})g_{(0)}^{kl}g_{(1)kl}
g_{(0)}^{ij}\partial_i\phi_{(0)}\partial_j\phi_{(0)} \nn
&& -{l \over 2}Y'(\phi_{(0)})\phi_{(1)}
{1 \over \sqrt{-g_{(0)}}}
\partial_i\left(\sqrt{-g_{(0)}}g_{(0)}^{ij}
\partial_j\phi_{(0)} \right) \nn
&& -{l \over 2}Y(\phi_{(0)}){1 \over \sqrt{-g_{(0)}}}
\partial_i\left(\sqrt{-g_{(0)}}g_{(0)}^{ij}\partial_j\phi_{(1)} \right) \\
&& \left. -{l \over 2}Y(\phi_{(0)}){1 \over \sqrt{-g_{(0)}}}
\partial_i\left\{\sqrt{-g_{(0)}}\left({1 \over 2}g_{(0)}^{kl}
g_{(1)kl}g_{(0)}^{ij}-g_{(0)}^{ik}g_{(0)}^{jl}
g_{(1)kl}\right) \partial_j\phi_{(0)} \right\}\right]
\ .\nonumber
\eea
Here we choose $\alpha$ to be $\alpha={6 \over l}$. Solving the 
equations of motion given by the variation of $S_{\rm ln}$ with respect 
to $g_{(1)ij}$ and $\phi_{(1)ij}$,  
\bea
\label{geq}
0&=&g^{ij}_{(0)}g^{kl}_{(0)}g_{(1)kl}-g^{ik}_{(0)}g^{jl}_{(0)}g_{(1)kl} \nn
&&+{l^2 \over 2}\left(R_{(0)}^{ij}-{1 \over 2}g_{(0)}^{ij}R_{(0)}\right) \nn
&&-{l^2 \over 2}V(\phi_{(0)})\left({1 \over 2}g^{ij}_{(0)}g^{kl}_{(0)}
-g^{ik}_{(0)}g^{jl}_{(0)}\right)\partial_k\phi_{(0)}\partial_l\phi_{(0)} \\
\label{peq}
0&=&{8 \over l^2}V(\phi_{(0)})\phi_{(1)}
+V'(\phi_{(0)})g^{ij}_{(0)}\partial_i\phi_{(0)}\partial_j\phi_{(0)} \nn
&& -{2 \over \sqrt{g_{(0)}}}\partial_i\left(
V(\phi_{(0)})\sqrt{g_{(0)}}g^{ij}_{(0)}\partial_j\phi_{(0)}\right)
\eea
$g_{(1)ij}$ and $\phi_{(1)ij}$ can be given in terms of $g_{(0)ij}$ 
and $\phi_{(0)}$:
\bea
\label{xiv}
g_{(1)ij}&=&{l^2 \over 2}\left( R_{(0)ij} - {1 \over 6}g_{(0)ij}R_{(0)}
\right) \nn
&& + {l^2 \over 2}V(\phi_{(0)})\left(\partial_i\phi_{(0)}
\partial_j\phi_{(0)} - {1 \over 6}g_{(0)ij}g_{(0)}^{kl}
\partial_k\phi_{(0)}\partial_l\phi_{(0)} \right) \\
\label{xv}
\phi_{(1)}&=&{l^2 \over 8}\left\{{V'(\phi_{(0)}) \over V(\phi_{(0)})}
g_{(0)}^{ij}\partial_i\phi_{(0)}\partial_j\phi_{(0)} 
+ 2 {1 \over \sqrt{-g_{(0)}}} \partial_i\left(\sqrt{-g_{(0)}}
g_{(0)}^{ij}\partial_j\phi_{(0)} \right)\right\} \ .
\eea
Here $V(\phi_{(0)})$ is defined in (\ref{xviii}). Substituting (\ref{xiv}) 
and (\ref{xv}) into (\ref{xiii}) and integrating it by parts, we find 
\bea
\label{xvii}
S_{\rm ln}&=&{l^3 \over 16\pi G}\int d^4x \sqrt{-g_{(0)}} 
\left[ {1 \over 8}R_{(0)ij}R_{(0)}^{ij}
-{1 \over 24}R_{(0)}^2 \right. \nn
&& + {V(\phi_{(0)}) \over 4} R_{(0)}^{ij}\partial_i\phi_{(0)}
\partial_j\phi_{(0)} - {V(\phi_{(0)}) \over 12} R_{(0)}g_{(0)}^{ij}
\partial_i\phi_{(0)}\partial_j\phi_{(0)}  \nn
&& + \left({V(\phi_{(0)})^2 \over 12} + {V'(\phi_{(0)})^2 \over 
32 V(\phi_{(0)})}\right)
\left(g_{(0)}^{ij}\partial_i\phi_{(0)}\partial_j\phi_{(0)} \right)^2 \nn
&& + {V'(\phi_{(0)}) \over 8} 
g_{(0)}^{kl}\partial_k\phi_{(0)}\partial_l\phi_{(0)} 
{1 \over \sqrt{-g_{(0)}}} \partial_i\left(\sqrt{-g_{(0)}}
g_{(0)}^{ij}\partial_j\phi_{(0)} \right) \nn
&& \left. + {V(\phi_{(0)}) \over 8}
\left\{{1 \over \sqrt{-g_{(0)}}} \partial_i\left(\sqrt{-g_{(0)}}
g_{(0)}^{ij}\partial_j\phi_{(0)} \right)\right\}^2 \right]\ .
\eea
Using the field $\varphi_{(0)}$ defined by 
\be
\label{xviiib}
\varphi_{(0)}\equiv\int d\phi_{(0)}\sqrt{2V(\phi_{(0)})}
\ee
in a similar way to (\ref{xviii}), we can rewrite (\ref{xvii}) as follows
\bea
\label{xix}
S_{\rm ln}&=&{l^3 \over 16\pi G}\int d^4x \sqrt{-g_{(0)}} 
\left[ {1 \over 8}R_{(0)ij}R_{(0)}^{ij}
-{1 \over 24}R_{(0)}^2 \right. \nn
&& + {1 \over 2} R_{(0)}^{ij}\partial_i\varphi_{(0)}
\partial_j\varphi_{(0)} - {1 \over 6} R_{(0)}g_{(0)}^{ij}
\partial_i\varphi_{(0)}\partial_j\varphi_{(0)}  \nn
&& \left. + {1 \over 4}
\left\{{1 \over \sqrt{-g_{(0)}}} \partial_i\left(\sqrt{-g_{(0)}}
g_{(0)}^{ij}\partial_j\varphi_{(0)} \right)\right\}^2 + {1 \over 3}
\left(g_{(0)}^{ij}\partial_i\varphi_{(0)}\partial_j\varphi_{(0)} 
\right)^2 \right]\ .
\eea
The Weyl anomaly coming from the multiplets of ${\cal N}=4$ 
supersymmetric $U(N)$ or $SU(N)$ Yang-Mills coupled with ${\cal N}=4$ 
conformal supergravity was calculated in \cite{HT}:\footnote{See Eqs.(2.5) 
and (2.6) in \cite{HT}.}
\bea
\label{xxi}
T&=&-{N^2 \over 4(4\pi)^2}\left[2\left(R_{ij}R^{ij}
-{1 \over 3}R^2\right)+F^{ij}F_{ij} \right. \\
&& \left. + 4\left\{ 2\left( R^{ij} - {1 \over 3} Rg^{ij}\right)
\partial_i\varphi^*\partial_j\varphi  
+\left|{1 \over \sqrt{-g}} \partial_i\left(\sqrt{-g}
g^{ij}\partial_j\varphi \right)\right|^2 \right\} + \cdots \right]\ .
\nonumber
\eea
Here $F_{ij}$ is the field strength of SU(4) gauge fields, $\varphi$ is 
a complex scalar field which is a combination of dilaton and RR scalar 
and ``$\cdots$'' expresses the terms containing other fields in 
${\cal N}=4$ conformal supergravity multiplet and higher powers of the 
fields.

If we choose
\be
\label{xx}
{l^3 \over 16\pi G}={2N^2 \over (4\pi)^2}\ ,
\ee
and consider the background where only gravity and the real part of the 
scalar field $\varphi$ in the ${\cal N}=4$ conformal supergravity 
multiplet are non-trivial and other fields vanish in (\ref{xxi}), Eq.(\ref{xix}) exactly reproduces the result in (\ref{xxi}) by 
using (\ref{vib}.

Hence, we got $d=2$ and $d=4$ holografic conformal anomaly for dilaton
coupled theories from supergravity side. The results of this study 
give further check of AdS/CFT correspondence in the presence of dilaton. 
It is also possible to extend this work and to find dilaton 
dependent conformal anomaly in higher dimensions, like $d=6,8$ etc. 
However, such calculation is very complicated as even in case of $d=6$ 
the number of invariants in conformal anomaly is rapidly growing. 
Nevertheless, using some classification of invariants (see, for 
example, \cite{DS,IO}) may significally simplify the result. Note that $d=6$ 
conformal anomaly for $(0,2)$ theory has been found in \cite{HS} very 
recently on purely gravitational background.

\end{document}